# Photon localization laser

Valery Milner and Azriel Z. Genack

*Department of Physics, Queens College of the City University of New York, Flushing, NY 11367, USA*

We demonstrate low-threshold random lasing in random amplifying layered medium via photon localization. Lasing is facilitated by resonant excitation of localized modes at the pump laser wavelength, which are peaked deep within the sample with greatly enhanced intensity. Emission occurs into long-lived localized modes overlapping the localized gain region. This mechanism overcomes a fundamental barrier to reducing lasing thresholds in diffusive random lasers, in which multiple scattering restricts the excitation region to the proximity of the sample surface.

In conventional lasers, emission is stimulated into well-defined cavity modes and emerges as a coherent beam. Care is taken to suppress scattering within the cavity since this would shorten the photon residence time in the lasing mode and thereby raise the excitation power required to initiate lasing. In the opposite limit of a random amplifying medium, however, multiple scattering impedes the flow of light out of the gain region [1,2]. This has led to a search for lasing action in random amplifying media, which was observed in rare earth [3] and semiconductor [4] powders, colloidal suspensions in dye solutions [5,6], and organic films [7]. However, the lasing threshold in these diffusive systems is not appreciably lowered below that for amplified spontaneous emission in homogeneous media because multiple scattering also impedes the flow of the incident pump light into the sample [8]. This creates a shallow gain region and allows subsequently emitted photons to escape promptly as diffuse luminescence.

In this Letter, we show that a collimated laser beam is produced at greatly reduced lasing threshold in a stack of microscope cover slides with interspersed dye films. The threshold is sufficiently low for lasing to be induced with a quasi-cw illumination by a 3-Watt laser beam from an Argon-ion laser. Pump energy is deposited deep within the sample, where it is greatly enhanced, by the resonant excitation of spectrally narrow localized modes. Subsequent emission occurs into long-lived localized modes which spatially overlap the region of molecular excitation. This is demonstrated by the strong correlation of pump transmission and output laser power.

The sample is a stack of $L$ partially reflecting glass slides of random thickness of ~100 μm with an amplifying region or an air gap between the slides. The amplifying layers consisted either of 25-μm thick solution of Rhodamine 6G laser dye in ethylene glycol, or of a single 0.3-mm-thick chromophore-doped plastic sheet. The stack was illuminated at normal incidence by a 150 ns pulse from a frequency-doubled, Q-switched Nd:YAG laser at 532 nm focused to a 100-μm spot.

Light in this one-dimensional (1D) geometry is localized by multiple scattering from the parallel layers which returns the wave upon itself [9-12]. Emission spectra, recorded with a 0.07-nm-resolution spectrometer, are shown in Fig. 1. The broad emission spectrum of neat dye solution (dashed curve in Fig. 1a) is compared with the emission spectrum from the random stack with interspersed dye layers slightly below and above the lasing threshold. Below threshold, the spectrum shows resolution-limited peaks of the electro-magnetic modes of the system (lower curve in Fig. 1a). Above threshold, a collimated emission beam perpendicular to the sample layers was observed (middle curve in Fig. 1a). The abrupt change in the output power with increasing pump energy shown in the inset to Fig. 1c is characteristic of the lasing transition. At low pump energies, lasing occurred in a single narrow line (Figs. 1a and 1b) while at higher energies, multimode lasing was observed (Fig. 1c) with wavelength and intensity that vary randomly with the position of the pump beam on the sample surface. The lasing threshold depended strongly upon the dye concentration, sample thickness, and location of gain inside the stack. The lowest threshold of ~ 0.25 μJ was measured with a single dye layer with a concentration of 4 mg of dye in 1 ml of ethylene glycol at the center of a stack of 60 glass slides. This pump energy is equivalent to a continuous pump intensity of ~ 10 kW/cm$^2$, which can be obtained from a standard continuous-wave laser source. Indeed, single and multimode lasing was observed using the output of a 6-Watt Argon-ion laser which was chopped in order not to overheat the sample (dashed curve in Fig. 1 b).

The distinct spontaneous emission and lasing peaks, shown in Fig.1, contrast with the continuous emission spectrum of the diffusive random laser [5] and is characteristic of localized electromagnetic modes. In localized media, the spectral width of a mode, δν, which is the inverse of the escape time of photons from the medium, is smaller than the typical spacing between modes, $\Delta\nu$, giving a value for the Thouless number, $\delta$, below unity, $\delta \equiv \delta\nu/\Delta\nu < 1$ [10,12,13]. Narrow emission modes have also been observed recently in micron-sized diffusive ZnO clusters [14] and organic thin films [7]. Although $\delta > 1$ in these samples, its value approaches unity when the photon excursion volume approaches the coherence volume $V_c = (\lambda/2)^3$. In the presence of substantial gain, emission into a few long-lived modes of the microscopic sample is enhanced, giving rise to distinct spectral peaks [15]. In contrast, the overlap of many modes in macroscopic diffusive lasers results in non-resonant feedback and a continuous emission spectrum [2]. Because of the large number of competing modes within the excitation volume, the number of stimulated photons in each mode is low. As a result, the lasing threshold is high and only a modest spectral narrowing is achieved.

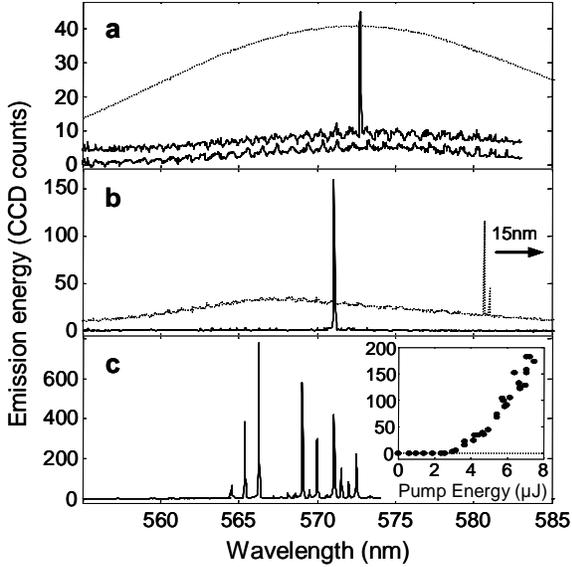

FIG. 1. Emission spectra of the neat dye solution (dashed curve in (a)) and of localization laser for the following experimental parameters: (a) – single 25-μm dye layer (2 mg of dye in 1 ml of ethylene glycol) in the middle of a stack of 50 glass slides; pump energy, $E_{pump}$, was 110 μJ and 130 μJ per pulse for the lower and middle curves (shifted up for clarity), respectively. (b,c) - sample of 100 layers, in which every other air gap was filled with the same dye solution; $E_{pump}$ = 3.2 μJ and 4.9 μJ per pulse for (b) and (c), respectively. The inset in (c) shows the total emission energy integrated over the spectral region 550-590 nm. Dashed curve in (b) shows the emission spectrum (shifted by 15 nm) obtained with a single colored plastic sheet in the center of a stack of 50 glass slides, which was pumped with a chopped 3-Watt cw beam from an Argon ion laser.

The small value of δ in random layered media reflects weak coupling between adjacent regions of a localized sample since modes in different regions of a sample do not spectrally overlap. This results in the exponential falloff of the average transmittance, $\langle T \rangle$, with sample thickness [9,10]. Excellent agreement is found in a comparison of the measured average transmittance over a 10mm² area of the sample and scattering matrix simulations [16] for random ensembles of 1D samples for $L \leq 60$ slides, shown in Fig. 2. The measured transmittance contained many realizations of longitudinal disorder as a result of the random speckle pattern in the transmitted field due to the transverse inhomogeneity of the glass slides of ~ 200nm/mm. A single exponential fit to the measured transmittance, $\langle T \rangle \sim \exp(-L/2\xi)$, for $L > 25$ slides gave the localization length $\xi = 13.5$ slides.

Even though the average optical transport is suppressed, resonant tunnelling through localized states provides spectrally narrow modes, in which transmission can be high [10,12,17]. A large number of narrow modes was observed in measurements (Fig. 3a) and in scattering matrix calculations (Fig. 3b) of transmittance for random stacks of 50 glass slides. The laser beam was focused to an 80-μm-diameter spot. Stronger focusing was avoided in order to prevent spectral line broadening due to the angular spread of the incident beam. High spectral resolution was obtained using a single-mode tunable dye laser with a linewidth of ~ 1 MHz ($\Delta\lambda \approx 10^{-6}$ nm). In the spectrum shown in Fig. 3a, the laser was scanned in 0.01-nm steps. Measurements of the width of a typical narrow isolated transmission line using scans with step size of $10^{-4}$ nm gave $\delta\lambda = 4.3\times10^{-3}$ nm. Narrow transmittance lines, which correspond to long-lived spatially localized modes [12,18], occur at both the excitation and emission wavelengths. The highest transmittance found in simulations is of order unity and occurs at the wavelength corresponding to the peaks of the narrowest lines at which the intensity is exponentially enhanced near the center of the layered medium, as shown in Figs. 3b and 3c. The spatial distribution of the pump energy at wavelength $\lambda_1$ inside the layered structure demonstrates the deep penetration of the pump energy into the sample's interior when the pump laser is tuned to resonance with a localized mode. Energy absorbed from this mode is subsequently emitted into long-lived localized modes which fall within the dye emission spectrum, e.g. at wavelength $\lambda_2$. Stimulated emission is enhanced when the spatial energy distributions at both the excitation and emission wavelengths overlap. In contrast to the resonant modes at $\lambda_1$ and $\lambda_2$, the energy of nonresonant light at wavelength $\lambda_3$ falls exponentially within the sample. The deposition of pump energy deep within the sample and its efficient coupling to long-lived emission modes removes a major barrier to achieving low-threshold lasing in the presence of disorder. Serendipitously, the increased penetration of the excitation light entails an exponential increase in the pump intensity near the center of the sample. This produces population inversion even at low incident power.

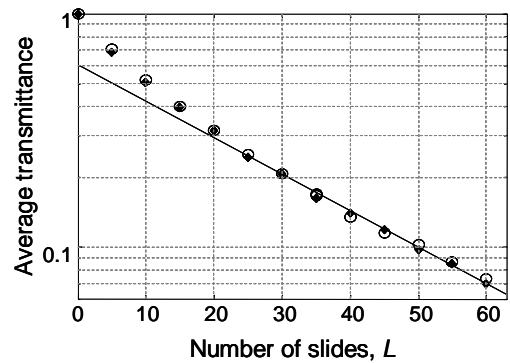

FIG. 2. Average transmittance through a stack of glass slides without intervening dye solution. Experimental results (circles) are compared with the results of numerical calculations (diamonds), in which the averaging was performed over 100 random realizations of the slide thickness.

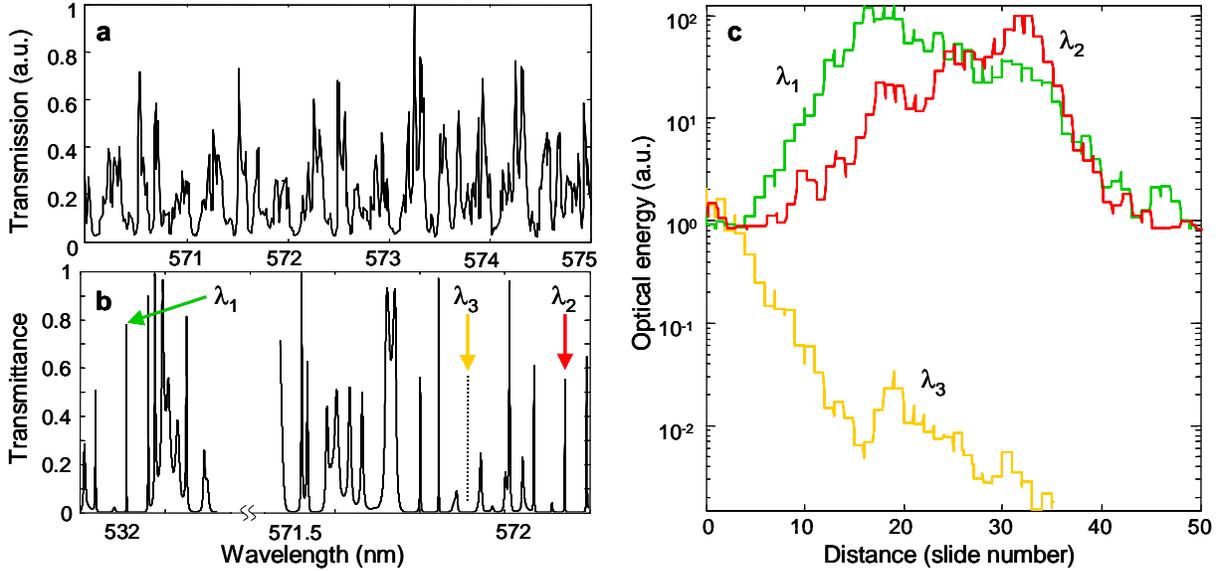

FIG. 3. (Color online) (a) High-resolution experimental spectrum of transmission through a stack of 50 glass slides. (b) Results of a scattering-matrix calculation of the transmittance spectrum for a stack of 50 slabs of random thickness illuminated by a plane wave. (c) Calculated spatial energy distribution of a plane wave incident upon the sample in (b) for the three wavelengths indicated. A stepwise change in energy density occurs at the air/glass interfaces.

Since lasing occurs via the excitation of narrow localized modes, large fluctuations in the output lasing power may be expected as different parts of the sample are excited [18]. This is seen in the plot of the total energy emitted into a solid angle of 0.2 steradian versus the transverse position of the pump beam on the sample surface in Fig. 4. Different realizations of disorder were sampled as different parts of the stack were excited with 10 μJ pulses while the sample was translated in the plane of the glass slides. The total laser power was integrated over the wavelength range of 555-590 nm. The simultaneously measured pump transmission (dashed curve) is strongly correlated with the output laser power. This is contrary to the expectation for a diffusive amplifying medium that strong transmission would be the result of weak absorption of the pump laser, and thus would be associated with low output laser power. In the present case, strong transmission is expected when the pump laser is at resonance with localized modes which are spatially peaked near the center of the sample. Pump intensity is therefore exponentially enhanced within the sample (Fig. 3), which results in efficient energy transfer to the gain medium. Further, since excitation in the center of the sample is likely to escape via emission into the localized modes with long photon residence time, the opportunity for stimulated emission is enhanced and the laser output power is high. Since both strong pump transmission and high output laser power are associated with the excitation of localized modes in the center of the sample, these quantities are strongly correlated.

The crossover from nonresonant to resonant feedback in random lasers as the parameter δ falls below unity parallels the transition between wave diffusion and localization in random media at the localization threshold of $\delta = 1$ [10]. The inverse of the Thouless number is the average number of returns of a wave to a coherence volume $V_c$ within a disordered sample [13]. Thus, the probability of a wave trajectory returning upon itself is small in a diffusive medium with $\delta > 1$. Wave interference may therefore be neglected in calculating average optical propagation. The evolution of light within such an amplifying diffusive medium may then be described as a random walk of photons at the pump and emission wavelengths, corresponding to nonresonant feedback [19]. In contrast, multiple returns to a coherence volume lead to photon localization and the condition $\delta < 1$, which is associated with resonant feedback in the random medium.

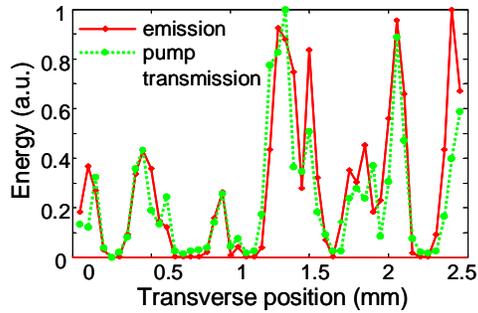

FIG. 4. (Color online) Spatial variation of transmitted energy of the pump beam with fixed input power of 10 μJ/pulse (dashed curve), and of the total laser energy (solid curve) for a stack of 50 glass slides with a single layer of dye solution in the center.

The photon localization laser described here is of simple design, which can be literally slapped together. A collimated beam emerges normal to the surface since lasing is initiated only in the longest-lived modes, which spatially overlap the gain region. Low-threshold, narrow-line lasing was observed in samples in which $L$ is not much larger than $\xi$. Since the mode lifetime is expected to grow exponentially with the number of layers in a 1D sample, greatly improved performance may be anticipated in samples with larger ratio $L/\xi$, which may be produced, for example, in binary thin films. When the transverse spread of a wave becomes comparable to the transverse coherence length set by inhomogeneity in the plane [16], the sample can no longer be regarded as one-dimensional and the role of transverse disorder must be considered.

We thank V. I. Kopp for use of the 1D computer simulation program, which he developed, and for valuable discussions. This work is sponsored by the US Army Research Office (DAAD190010362) and the National Science Foundation (DMR0205186).